# Spatio-spectral classification of hyperspectral images for brain cancer detection during surgical operations

Himar Fabelo[1]*, Samuel Ortega[1], Daniele Ravi[2], B. Ravi Kiran[3], Coralia Sosa[4], Diederik Bulters[5], Gustavo M. Callicó[1], Harry Bulstrode[6], Adam Szolna[4], Juan F. Piñeiro[4], Silvester Kabwama[5]‡, Daniel Madroñal[7]‡, Raquel Lazcano[7]‡, Aruma J-O'Shanahan[4]‡, Sara Bisshopp[4]‡, María Hernández[4]‡, Abelardo Báez[1], Guang-Zhong Yang[2], Bogdan Stanciulescu[8], Rubén Salvador[7], Eduardo Juárez[7], Roberto Sarmiento[1]

**1** Institute for Applied Microelectronics (IUMA), University of Las Palmas de Gran Canaria (ULPGC), Las Palmas de Gran Canaria, Spain, **2** The Hamlyn Centre, Imperial College London (ICL), London, United Kingdom, **3** Laboratoire CRISTAL, Université Lille 3, Villeneuve-d'Ascq, France, **4** Department of Neurosurgery, University Hospital Doctor Negrin, Las Palmas de Gran Canaria, Spain, **5** Wessex Neurological Centre, University Hospital Southampton, Tremona Road, Southampton, United Kingdom, **6** Department of Neurosurgery, Addenbrookes Hospital, University of Cambridge, Cambridge, United Kingdom, **7** Centre of Software Technologies and Multimedia Systems (CITSEM), Universidad Politecnica de Madrid (UPM), Madrid, Spain, **8** Ecole Nationale Supérieure des Mines de Paris (ENSMP), MINES ParisTech, Paris, France

☯ These authors contributed equally to this work.
‡ These authors also contributed equally to this work.
* hfabelo@iuma.ulpgc.es

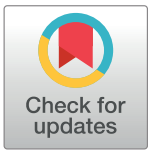

## Abstract

Surgery for brain cancer is a major problem in neurosurgery. The diffuse infiltration into the surrounding normal brain by these tumors makes their accurate identification by the naked eye difficult. Since surgery is the common treatment for brain cancer, an accurate radical resection of the tumor leads to improved survival rates for patients. However, the identification of the tumor boundaries during surgery is challenging. Hyperspectral imaging is a non-contact, non-ionizing and non-invasive technique suitable for medical diagnosis. This study presents the development of a novel classification method taking into account the spatial and spectral characteristics of the hyperspectral images to help neurosurgeons to accurately determine the tumor boundaries in surgical-time during the resection, avoiding excessive excision of normal tissue or unintentionally leaving residual tumor. The algorithm proposed in this study to approach an efficient solution consists of a hybrid framework that combines both supervised and unsupervised machine learning methods. Firstly, a supervised pixel-wise classification using a Support Vector Machine classifier is performed. The generated classification map is spatially homogenized using a one-band representation of the HS cube, employing the Fixed Reference t-Stochastic Neighbors Embedding dimensional reduction algorithm, and performing a K-Nearest Neighbors filtering. The information generated by the supervised stage is combined with a segmentation map obtained via unsupervised clustering employing a Hierarchical K-Means algorithm. The fusion is performed using a majority voting approach that associates each cluster with a certain class. To evaluate the proposed approach, five hyperspectral images of surface of the brain affected by glioblastoma tumor in vivo from five different patients have been used. The final classification maps obtained have been analyzed and validated by specialists. These preliminary results are promising, obtaining an accurate delineation of the tumor area.

## Introduction

In addition to radiotherapy and chemotherapy, surgery is one of the major treatment options for brain tumors [1]. However, because brain tumors infiltrate and diffuse into the surrounding normal brain, the surgeon's naked eye is often unable to accurately distinguish between the tumor and normal brain tissue. Inevitably, tumor tissue is either unintentionally left behind during surgery or too much normal brain tissue is taken out. Studies have shown that tumor tissue left behind during surgery is the most common cause of tumor recurrence and is a major cause of morbidity and mortality [2–4]. On the other hand, over-resection of brain tumor tissues has also been shown to cause permanent neurological deficits that affect patients' quality of life [5]. Intra-operative neuro-navigation, intra-operative Magnetic Resonance Imaging (iMRI) and fluorescent tumor markers such as 5-aminolevulinic acid (5-ALA) have been developed as adjuncts to surgery to help with brain tumor delineation. Although these adjuncts have improved the accuracy of brain tumor resections, they have a number of limitations. Neuro-navigation is rendered inaccurate at locating tumor margins due to brain shift and changes in tumor volume that occurs during resections [6,7]. Intra-operative Magnetic Resonance Imaging was developed as a solution to intra-operative brain shift capable tumor margin mapping intra-operatively. However, this has been found to have poor spatial resolution, to largely extend the surgery time and it is very expensive [8]. Due to the time to stop surgery and obtain scans it is better regarded as providing at most a few images at certain timepoints rather than a continuous real time image. Fluorescent tumor markers such as 5-aminolevulinic acid (5-ALA) are excellent at identifying tumors but can only be used for high grade tumors, produce important knock-on effects and are poor at defining tumor margins mainly due to the diffuse nature of brain tumors [9,10].

Therefore, despite the improvement in surgery and technology, we are still unable to accurately define brain tumor margins. Label free, non-ionizing imaging modalities that rely on intrinsic properties of tumors or normal brain could be a potential solution to the above problem. Hyperspectral Imaging (HSI) is a form of imaging spectroscopy that captures spectral and spatial data beyond the limited three electromagnetic bands of the human eye [11]. It produces a three-dimensional image with each pixel containing spectral information of the captured scene. The spectral information of each pixel correlates to the chemical composition of the scene. This technology has relevance in the medical field because it has been proven that the interaction between electronic radiation and tissue carries useful information for diagnosis purposes [12]. In the field of early detection of tumor, HSI is shown as a promising technology due to its non-invasive interaction with tissue and its capability to rapidly acquire and analyze data, obtaining useful information for diagnosis purposes. In recent years, the number of studies using HSI analysis for cancer diagnosis has rapidly increased. The main differences between studies are in the acquisition system setup as in [13], the nature of the samples (in-vivo, ex-vivo or in-vitro) the disease studied (prostate [14], ovaries [15], breast [16], tongue cancer [17], skin and lung cancer [18] or oral cancer [19]), and the applied processing methods to analyze the HS data (as in the application to larynx cancer [20]).

One of the most active research groups in biomedical applications of HSI is led by Professor Baowei Fei, from the Department of Biomedical Engineer at Emory University. The main characteristics of the HSI research performed by this group can be summarized as follows. Their experiments explore cancer diseases in animal subjects. Until now, they have analyzed prostate cancer [21] and head and neck cancer [22]. Moreover, they usually work using an acquisition system based on LCTFs (Liquid Crystal Tunable Filters) in the VNIR spectral range, from 400 nm to 1000 nm. Most of their experiments have been carried out in-vivo during surgical procedures. Their research has exhaustively analyzed which pre-processing techniques are more suitable to compensate the variations of the environmental conditions during the acquisition inside an operating theatre [23,24]. The processing techniques employed by this research group in order to extract useful information from the hyperspectral (HS) cubes, vary depending on each research study, but each new publication presents novel and sophisticated methods such as a Minimum-Spanning Forest classification [25].

In this pilot study, we investigate whether intra-operative hyperspectral imaging can identify and delineate brain tumors. This work is framed in a European collaborative project funded by the Research Executive Agency (REA) called HELICoiD (HyperEspectraL Imaging Cancer Detection) formed by four universities, two university hospitals and three leading industry partners.

## Materials and methods

### Intra-operative hyperspectral acquisition system

The hyperspectral acquisition system employed in this work is called the HELICoiD demonstrator [26]. The system is composed by a hyperspectral pushbroom camera manufactured by HeadWall Photonics: the Hyperspec® VNIR A-Series model. The VNIR camera covers the spectral range from 400 nm to 1000 nm, with a spectral resolution of 2–3 nm, being able to capture 826 spectral bands and 1004 spatial pixels. This device integrates a Silicon CCD detector array with a minimum frame rate of 90 fps, understanding in this context a frame as a line of 1004 pixels and 826 spectral bands. The lens used in this camera is a Xenoplan 1.4 with 22.5 mm of focal length and a broadband coating for the spectral range of 400 nm to 1000 nm. The camera is attached in a scanning platform composed by a stepper motor and a spindle capable of covering an effective area of 230 mm. This scanning platform is required to acquire the second spatial dimension as the pushbroom camera can only sample a single line. The setup uses an illumination system composed by a Quartz-Tungsten-Halogen (QTH) lamp connected to a cold light emitter via fiber optic that allows achieving cold illumination over the brain surface. This is required in order to avoid high temperatures produced by the QTH lamp over the brain surface. Fig 1 shows the intra-operative hyperspectral acquisition system being used during a neurosurgical operation.

### In-vivo human brain hyperspectral image database

A set of five in-vivo brain surface HS images, captured using the previously described acquisition system, has been used for this research. These images belong to adult patients undergoing craniotomy for resection of intra-axial brain tumor. Images have been obtained at the University Hospital Doctor Negrin of Las Palmas de Gran Canaria (Spain) and at the University Hospital of Southampton (United Kingdom) from five different patients with confirmed grade IV glioblastoma tumor on histopathology. The study protocol and consent procedures were approved by the Comité Ético de Investigación Clínica-Comité de Ética en la Investigación (CEIC/CEI) for the University Hospital Doctor Negrin and the National Research Ethics Service (NRES) Committee South Central—Oxford C for the University Hospital of Southampton.

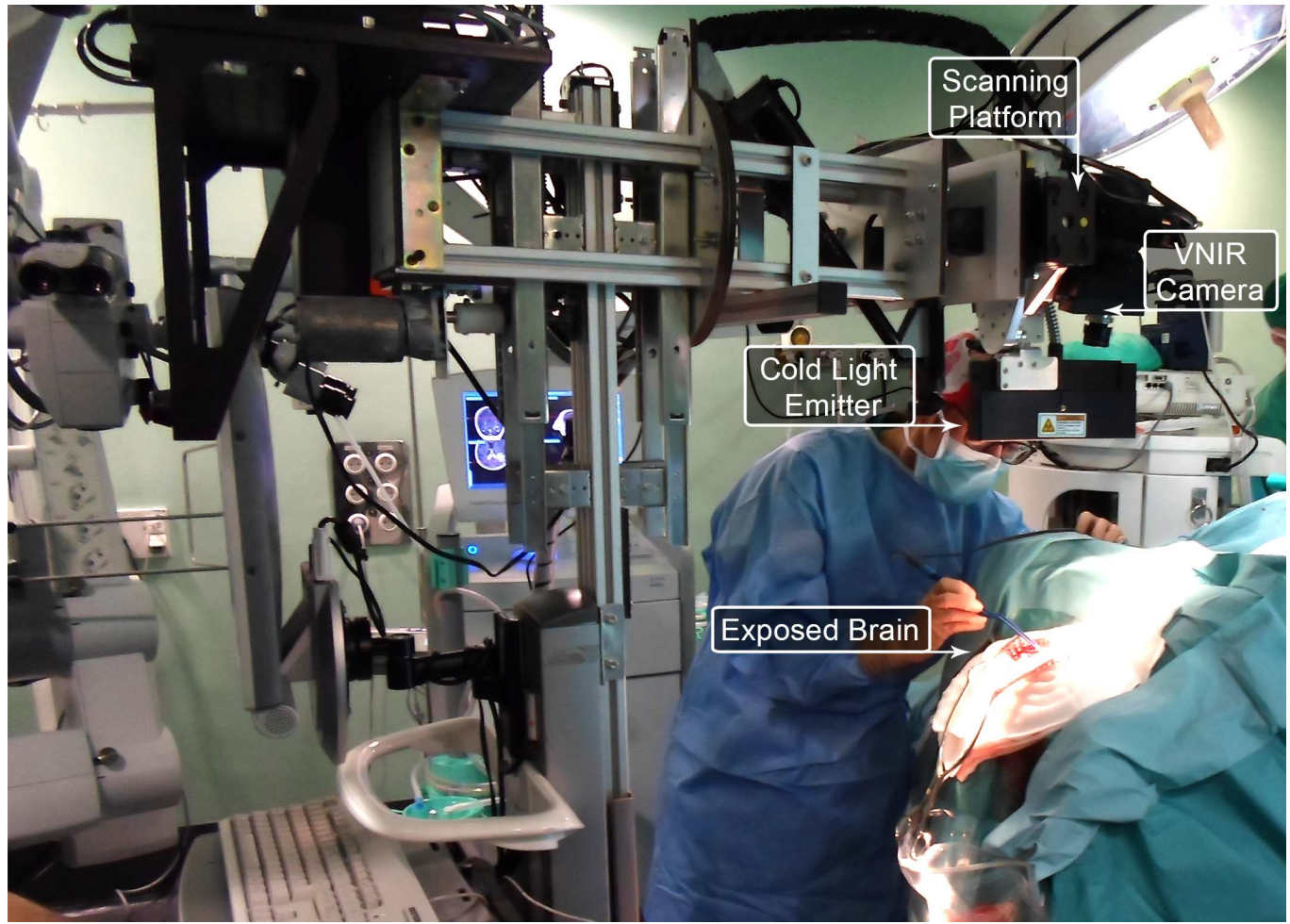

**Fig 1. Intra-operative hyperspectral acquisition system used during a neurosurgical procedure at the University Hospital Doctor Negrin of Las Palmas de Gran Canaria.**

https://doi.org/10.1371/journal.pone.0193721.g001

Written informed consent was obtained from all subjects. The individual that appears in Fig 1 in this manuscript has given written informed consent (as outlined in PLOS consent form) to publish these case details.

The procedure to acquire the in-vivo data during neurosurgical operations has been described elsewhere and in summary is as follows. First, after performing the craniotomy and durotomy, the neurosurgeons place some rubber ring markers over the brain surface where they are confident that the tissue inside the markers is tumor or normal based on its macroscopic appearance and taking into account the information provided by the neuronavigator from an MRI scan (Magnetic Resonance Image). In case of patient 1, two markers were placed in the tumor area and one marker was placed over the healthy tissue. In cases of patients 2, 3 and 4, two markers were use, one placed over the tumor area and another one placed over the normal tissue. Finally, in case of patient 5, no markers were used since the tumor was in a deeper layer with respect to the normal tissue and it was clearly identified. After that, the operator of the acquisition system captures a HS image. Depending on the location of the tumor, the images are acquired at various stages of the operation. In cases of patients 1, 2, 3 and 4, one image was obtained immediately after the dura was removed since the tumor was superficially located. In case of patient 5, the image was captured in an advanced stage of the tumor resection since the tumor was deep seated. Once the HS image is taken, the operating surgeon performs a biopsy of the tissue located within the tumor tissue marker/s (in case of patient 1, 2, 3 and 4) or within the clearly identified tumor area (in case of patient 5). The resected tissue is sent to the pathologists in order to confirm the presence or absence of tumor, and obtain the specific histopathological diagnosis (grade and type of tumor). Since this technology cannot penetrate into the tissue (in case of near infrared it can be 1 mm at most), the average size of the resected sample of the tumor for pathological analysis is 0.5x0.5 mm and 0.2 mm depth. Normal tissue markers are only used as a reference for the labelling process performed after the operation has finished. It is not ethic to perform a biopsy of the tissue that is known to belong to normal brain (it can produce damages in the undergoing patient outcomes). Employing the histopathological information (from the tumor tissue samples) and the knowledge of the operating surgeon (from the normal tissue samples), the labeling of the HS cubes is performed to generate a gold standard dataset for the supervised classification stage of the brain cancer detection algorithm.

The manual labeling of the HS data consists of visual identification of each sample by a specialist, which is time-consuming task and can introduce errors due to human intervention. For this reason, a methodology for extracting the gold standard information from the HS cubes, based on the Spectral Angle Mapper (SAM) algorithm, has been developed. The tool developed for sample labeling has been designed using Matlab® GUIDE application and measures the angle between two high dimensional vectors. This SAM classification is an automated method for comparing the spectra of the pixels of a HS image with a well-known spectrum obtained from a reference pixel. The tool was used by the corresponding operating surgeon after the operation conclusion in order to generate the gold standard map for each image. Four different classes were employed in this study: *normal tissue*, *tumor tissue*, *blood vessel* and *background*. The procedure to obtain the gold standard map is as follows. First, the user loads a HS cube to be labeled and selects a reference pixel, looking the synthetic RGB representation of the HS cube, at the location where a biopsy is done (where the tumor marker is placed) or at a location far enough from the tumor margins where the surgeon can be quite confident that the tissue is abnormal (in the case of tumor labeling). In case of normal tissue, blood vessel and background classes, the labeling is performed by the operating surgeon by selecting a reference pixel by naked eye based on their knowledge and experience. Then, the most similar pixels to the selected reference pixel are highlighted, based on the SAM measurement, and the user configures the threshold that varies the tolerances on the pixels selected. Once the user considers that only the pixels belonging to one class are highlighted, the selected pixels are assigned to that class. Fig 2 shows a screenshot of the HELICoiD Labelling Tool where the labeling procedure of the blood vessel class has been done. On the left side of the image (Fig 2A), the synthetic RGB representation of the HS cube is shown. In the center (Fig 2B), the SAM representation is presented, where only the pixels that have a spectral angle less than 0.08˚ respect to the selected reference pixel are highlighted. In this case, the reference pixel and its correspondent SAM representation belongs to the blood vessel class. Finally, on the right side of the image (Fig 2C), the gold standard map generated for patient 2 is shown, where tumor tissue, normal tissue, blood vessels and background are represented in red, green, blue and black colors respectively. Some sliders controls are presented in the labeling tool so as to adjust the gamma of the synthetic RGB image, the overlapping transparency of the SAM image over the synthetic RGB image and the threshold value. Employing this labeling tool, a total of 44,555 spectral signatures was labeled. Table 1 summarizes the total number of labeled spectral signatures generated for each class, the number of tumor biopsies performed and the number of images captured for each patient. Summarizing, the reliability of the gold standard

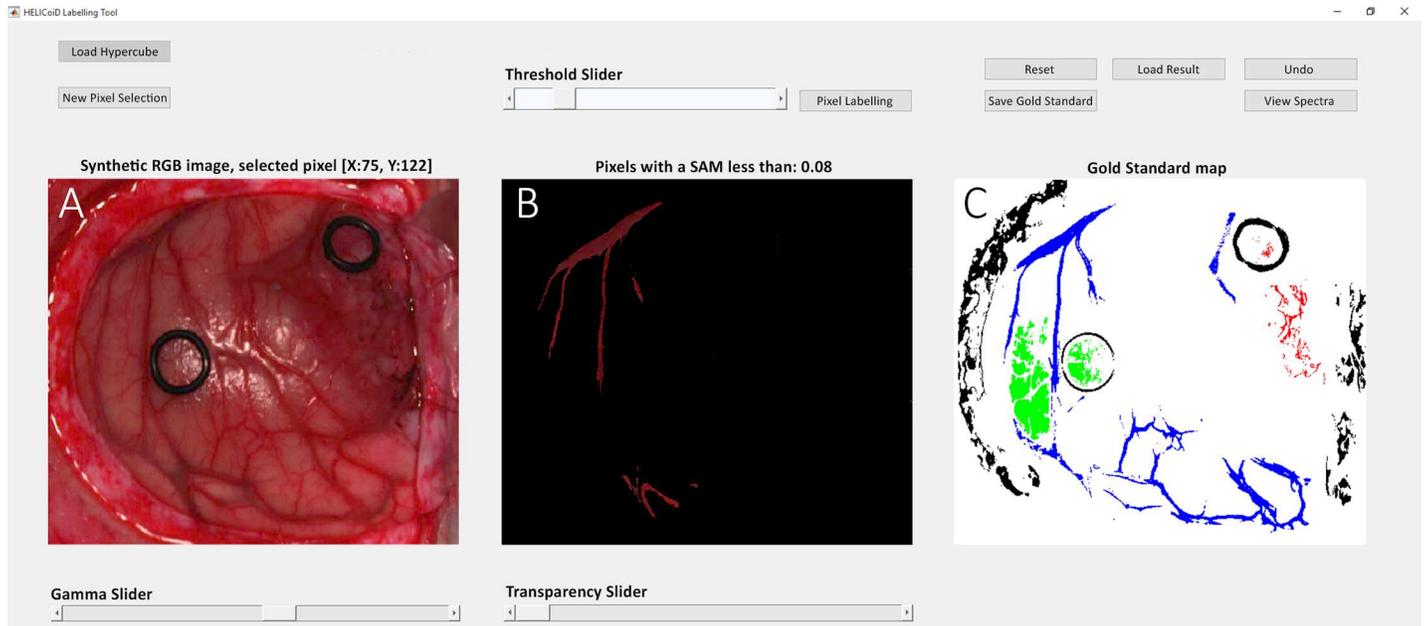

**Fig 2. Screenshot of the HELICoiD Labeling Tool.**

https://doi.org/10.1371/journal.pone.0193721.g002

is guaranteed by the use of the intraoperative MRI Neuronavigation for placing the rubber ring markers; the operating surgeon knowledge and experience for the labelling of the normal tissues, blood vessels and background samples; and finally, the pathological analysis of the resected tissue for the tumor labeling.

### Brain cancer detection algorithm

The classification framework developed in this study aims to exploit both the spatial and spectral features of the HS images. Fig 3 illustrates the scheme of this classification framework based on five main steps: data pre-processing, dimensional reduction, spatial-spectral supervised classification, unsupervised clustering segmentation and hybrid classification. After capturing the in-vivo brain surface HS cube (Fig 3A), the raw image is pre-processed in order to homogenize the spectral signatures of each pixel (Fig 3B). After the pre-processed stage, the golden standard employed for building the supervised classifier model is extracted by the specialists (Fig 3C) using the previously described labeling tool and the SVM classifier is trained (Fig 3D). Once the SVM model is generated, it is used to perform the supervised pixel-wise classification over the pre-processed HS cube (Fig 3E). Then, a spatial-spectral homogenization is accomplished [27] using a KNN (K-Nearest Neighbor) filtering (Fig 3G), where a one-

**Table 1. Gold standard dataset for the supervised training process.**

| Patient ID | #Captured Images | #Tumor Biopsies | Tissue Type (#pixels) | | | | Total (#pixels) |
|---|---|---|---|---|---|---|---|
| | | | Normal Tissue | Tumor Tissue | Blood Vessel | Background | |
| 1 | 1 | 2 | 2,295 | 1,221 | 1,331 | 630 | 5,477 |
| 2 | 1 | 1 | 4,516 | 855 | 8,697 | 1,685 | 15,753 |
| 3 | 1 | 1 | 1,251 | 2,046 | 4,089 | 696 | 8,082 |
| 4 | 1 | 1 | 1,842 | 3,655 | 1,513 | 2,625 | 9,635 |
| 5 | 1 | 1 | 977 | 1,221 | 907 | 2,503 | 5,608 |
| | | Total (#pixels) | 10,881 | 8,998 | 16,537 | 8,139 | 44,555 |

https://doi.org/10.1371/journal.pone.0193721.t001

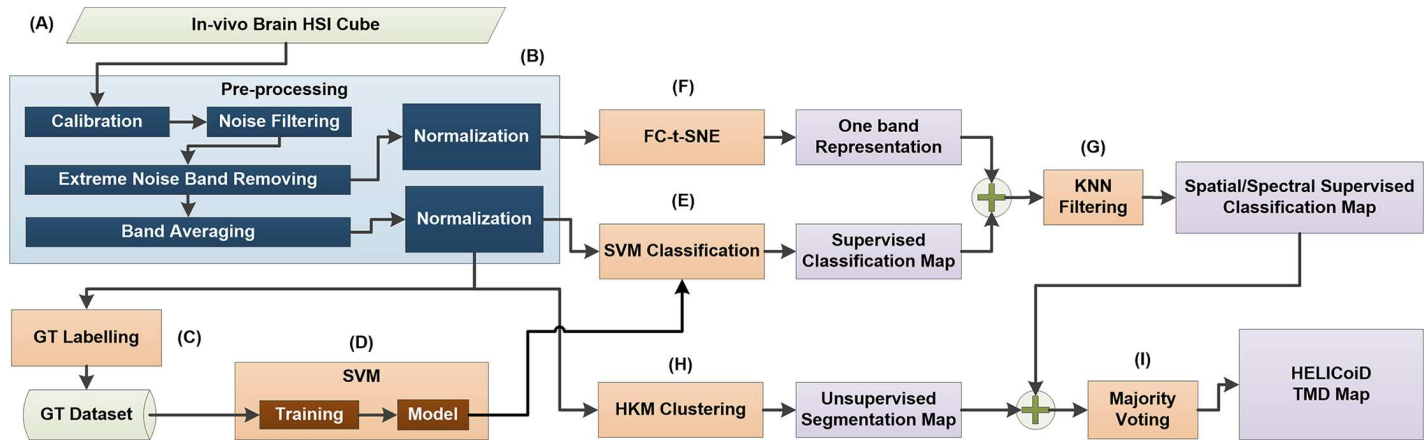

**Fig 3. Brain cancer detection and delimitation algorithm overview diagram.** (A) HS cube of in-vivo brain surface. (B) Pre-processing stage of the algorithm. (C) Database of labeling samples generation. (D) SVM model training process employing the labeled samples dataset. (E), (F) and (G) Algorithms that conform the spatial-spectral supervised classification stage. (H) and (I) Algorithms that generate the unsupervised segmentation map and the final HELICoiD TMD map, respectively.

https://doi.org/10.1371/journal.pone.0193721.g003

band representation of the HS cube is employed. The dimensionality reduction algorithm used to obtain the one-band representation of the HS cube is the FR-t-SNE algorithm (Fig 3F). This algorithm has been selected because it provides the best score along different HS images compared to other dimensionality reduction algorithms [28]. Once the spatial-spectral homogenization has been performed, a filtered classification map is available. In order to obtain the final classification map, the spatial-spectral supervised classification map is combined with a segmentation map obtained via unsupervised hierarchical clustering (Fig 3H) using a Majority Voting (MV) approach [29] (Fig 3I).

**Data pre-processing.** After the acquisition of the in-vivo brain surface HS cube (Fig 3A), a pre-processing chain, already explained in [30], is applied to the HS cube to homogenize the spectral signatures of each pixel (Fig 3B) and to reduce the dimensionality of the HS image without losing the main spectral information contained on it. This pre-processing chain consists of five steps. The first step performs a radiometric calibration of the raw spectral signature of each pixel using the black and white reference images acquired by the acquisition system inside the operating theatre with the same illumination conditions that the image that will be captured. The white reference image is obtained using a standard white reference tile and the dark reference image is acquired by keeping the camera shutter closed. Fig 4A and 4B show an example of a single raw spectral signature and the calibrated spectral signature of a grade IV glioblastoma tumor respectively. The second step applies noise filtering using the first stage of the HySIME algorithm where a function called Hyperspectral Noise Estimation infers the noise in the HS data, by assuming that the reflectance at a given band is well modeled by a linear regression on the remaining bands. Fig 4C plots the spectral signature after the HySIME noise filtering application. In the third step, the spectral bands from the lowest and highest bands are removed due to their low SNR because of the limited performance of the CCD sensor in these ranges. Bands from 0 to 50 and from 750 to 826 are removed. After the extreme noise band removing step, the spectral signatures are reduced in bands through spectral averaging due to the information redundancy between contiguous bands. The reduced HS cube is formed of 129 spectral bands. Finally, the last step of the pre-processing chain applies normalization over the samples to avoid the different radiation intensities of each pixel produced by the non-uniform surface of the brain. Fig 4D illustrates the final pre-processed spectral signature.

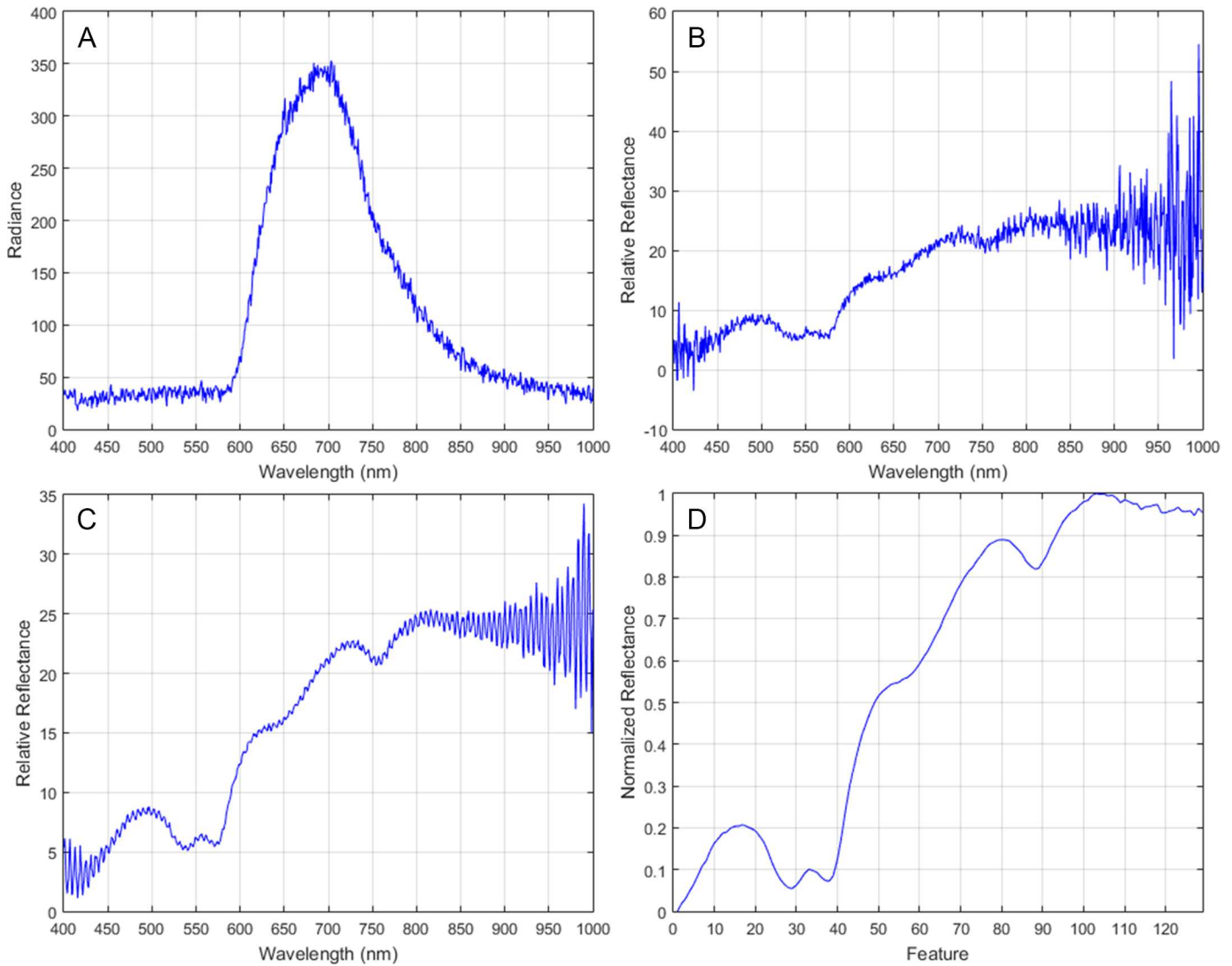

**Fig 4. Spectral signature of a grade IV glioblastoma tumor tissue.** (A) Raw spectral signature. (B) Calibrated spectral signature. (C) HySIME filtered spectral signature. (D) Final pre-processed spectral signature.

https://doi.org/10.1371/journal.pone.0193721.g004

**Dimensional reduction.** From an information-processing point of view, the intrinsic dimensionality of HS images can be significantly reduced before subsequent image characterization steps are applied. Dimensionality reduction maps high-dimensional data into a meaningful representation of reduced dimensional space so that the observed properties of the initial data are still preserved in the low dimensional space. Since the intrinsic dimension, as well as the geometry of the initial data, is unknown, dimensionality reduction, in general, is an ill-posed problem that can only be solved by assuming certain data properties.

Thus far, many algorithms for dimensionality reduction have been developed in literature [31]. Principal Component Analysis (PCA) [32] is one of the most popular linear techniques for dimensionality reduction. It maps the data preserving as much as possible their variance. However, PCA has two important limitations: it is based on a global property–the variance of the data–and it is a linear technique. Non-linear methods have the advantage that can deal better with complex real world data. Techniques such as Isomap [33], Locally Linear Embedding (LLE) [34], Hessian [35] and Laplacian [36] are examples of non-linear methods. In this paper, the Fixed Reference t-Distributed Stochastic Neighbors Embedding (FR-t-SNE) algorithm proposed in [28] is used as dimensional reduction for the HS images.

FR-t-SNE is an extension of the t-Distributed Stochastic Neighbors Embedding (t-SNE) [37] that is a nonlinear technique well suited for embedding high-dimensional data into a low dimensional space. As stated in [28], embedding a HS image using t-SNE may not guarantee consistent results since, at each dimensional reduction process of a new image, the random nature of the t-SNE can create embedded representations that are not persistent. Therefore, it can happen that similar tissues will be represented with different low dimensional representations across different images. This makes subsequent tissue characterization difficult. This problem is mainly generated by the lack of a fixed coordinate system, which does not allow the comparison of the embedded results across different tissue samples [38], FR-t-SNE tries to overcome these limitations by using a learning process aimed at finding a fixed reference coordinate system. FR-t-SNE is divided in three main steps: in Step 1, an optimal reference system is fixed to maintain a consistent manifold embedding along with all the images and circumvent the lack of a fixed coordinate system. In Step 2, the manifold is gradually tested on the training set using the predefined fixed reference. Finally, in the last step, a HS image is embedded efficiently. A KNN classification algorithm is used to obtain the low vector representation of each high dimensional vector after all the training images are processed and the manifold discovered. This KNN classifier will use a lookup table, containing the values of the learned reference coordinates to predict the embedded value of each sample in each new HS image.

In the proposed brain cancer detection algorithm, FR-t-SNE is employed to obtain a one-band representation of the pre-processed HS cube with 750 bands (without applying the band averaging step in the pre-processing chain).

**Spatial-spectral supervised classification.** Support Vector Machines (SVMs) are kernel-based supervised algorithms that have been extensively used for classification tasks. As a relevant example, a variant of the SVM classifier, called Fuzzy SVM classifier, was employed in the development of an emotion recognition system based on facial expression images, obtaining overall accuracy results of 96.77±0.10% [39]. In the HSI field, SVMs provide good performance for classifying this type of data when a limited number of training samples are available [40]. Due to its strong theoretical foundations, good generalization capabilities, low sensitivity to the problem of dimensionality and the ability to find optimal solutions, SVMs are usually selected by many researchers over other classification paradigms for classifying HS images [12]. In the medical field, SVMs have been used to detect multiple sclerosis subjects employing stationary wavelet entropy to extract features from magnetic resonance images used as input of the SVM classifier [41]. Furthermore, the same technique combined with a directed acyclic graph method has been used to diagnose unilateral hearing loss in structural MRI [42], demonstrating that the SVM algorithm is a reliable candidate to work with medical images. In medical HSI, SVMs have been already used to classify several types of cancer, including prostate [14], lung tissue and lymph nodes [21] skin tumors [43,44], tongue [45] and colon [46]. On the other hand, during the development of this research project, some studies have been carried out using SVMs to classify hyperspectral in-vivo images of human brain affected by cancer [26,30]. For the research presented in this paper, LIBSVM [47] has been used for support vector classification.

SVM algorithm requires a confident labeled dataset in order to train the model that will be used to classify the input data. In this work, the labeled dataset of in-vivo brain samples that is used to train the SVMs has been created by combining the efforts of neurosurgeons and pathologist, as it has been previously described. Before explaining the methodology employed for performing a supervised classification over the available HS data, some considerations have to be taken into account. Due to the impossibility of having a way to extract the labeled information from all the pixels in a HS cube of brain tissue, there are two ways of measuring the performance of the generated supervised models. For the available labeled dataset, it is possible to use standard metrics in order to measure the accuracy provided by the model when classifying unseen data. Nevertheless, for evaluating a supervised model applied to a whole HS cube (where not all pixels have been labeled) only the visual evaluation of an expert is possible. The methodology for evaluating the supervised classifiers in a quantitative way is as follows: first, we use the labeled information corresponding to the dataset, and then we apply a 10-fold cross validation in order to measure the performance of the model. The quantitative evaluation metrics used for this purpose has been sensitivity, specificity and overall accuracy metrics, and will be defined later in this paper.

Once the quantitative metrics have been obtained, the previously trained SVM classifier is used to classify a whole HS cube, and then it is evaluated by neurosurgeons in order to analyze the quality of the algorithm in distinguishing different types of tissues, materials or substances. In order to include the spatial features of the HS images, a spatial homogenization is applied to improve the supervised classification results by incorporating the neighborhood information of each pixel into the classification chain. The algorithm proposed in [27], which refines the pixel-wise classification probability map using a KNN filtering on non-local neighborhoods of a pixel, has been used. The algorithm has shown competitive classification accuracy results compared with other state-of-art spatial-spectral classification approaches [27]. The algorithm requires two inputs: the probability maps or confidence scores obtained from the supervised classifier ($P$) and the guidance image ($I$) (which is usually a one-band representation of the input HS image). The spatial-spectral feature vector is defined in Eq 1, where $I$ is the normalized pixel value (spectrum) at location $i$ and $l(i)$, $h(i)$ are the normalized longitude and latitude of the pixel $i$. The output of the KNN-filtering is given by Eq 2, where $N_i$ refers to the K-nearest neighbors of the pixel $i$ found in the feature space $F(i)$. It can be seen that at $\lambda = 0$ there is no spatial information, while when non-zero it captures the spatial information of pixel $i$ given by $l(i)$ and $h(i)$.

$$F(i) = (I(i), \lambda l(i), \lambda h(i)) \quad (1)$$

$$O(i) = \frac{\sum P(j)}{K}, j \in N_i \quad (2)$$

When $\lambda$ is set to zero, the spatial coordinates are not considered in the KNN filtering process, and when the value of $\lambda$ increases, the classification results tend to be oversmoothed, decreasing the accuracy of the classification results. The parameter $K$ has a similar influence in the classification results: when the $K$ value is high, the filtering method oversmooths the classification results, worsening the accuracy of the classification results. In this approach, it is not possible to provide a quantitative measure of the influence of $K$ and $\lambda$ parameters, due to the absence of a complete golden standard map. Nevertheless, the influence of these parameters in the generation of the classification maps has been studied. As mentioned before, large values of $K$ or $\lambda$ tends to oversmooth the obtained classification maps. Several executions of the KNN filtering were performed employing different values of $K$ (5, 10, 20, 40 and 60) and $\lambda$ (0, 1, 5, 10 and 100). Fig 5 shows the filtered classification maps of the patient 2 using different values of $K$ and $\lambda$. In both cases, small values of $K$ and $\lambda$ result in a mix of small classes that do not represent the real distribution of the tissues. On the other hand, large values of $K$ and $\lambda$ tend to oversmooth the classes. After a visual inspection of the results by the specialists (neurosurgeons), the final values of $K$ and $\lambda$ chosen for this study were $K = 40$ and $\lambda = 1$. These values

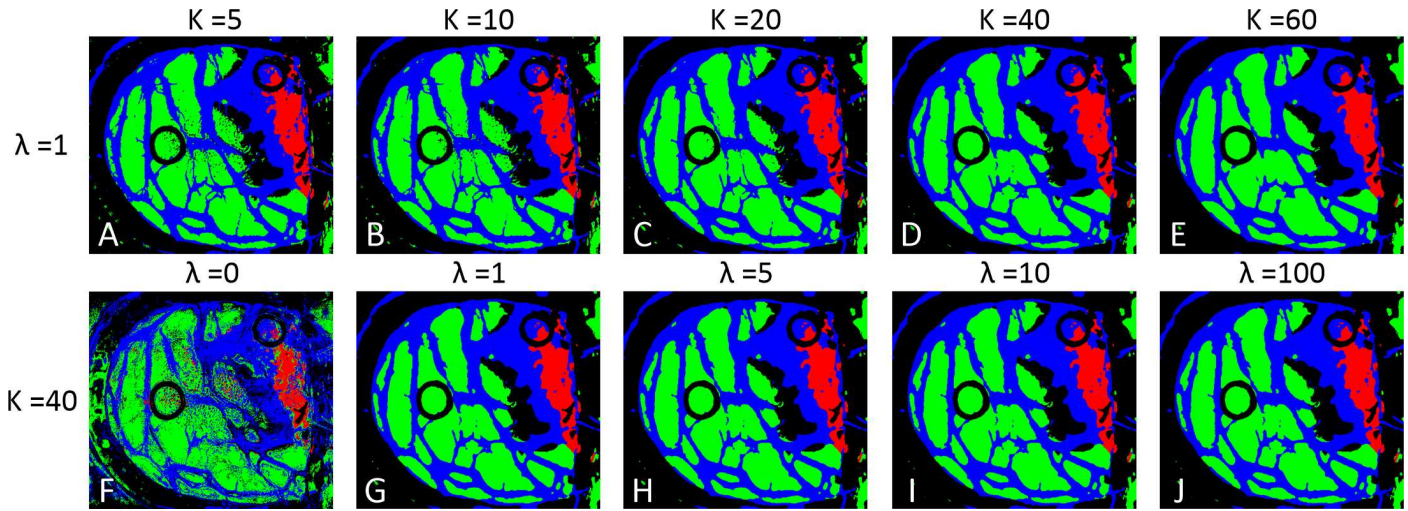

**Fig 5. KNN filtered maps obtained with different K and λ values.** (A), (B), (C), (D) and (E) filtered maps obtained with K equal to 5, 10, 20, 40, and 60, while keeping λ value fixed to 1. (F), (G), (H), (I) and (J) filtered maps obtained with λ equal to 0, 1, 5, 10, and 100, while keeping K value fixed to 40.

https://doi.org/10.1371/journal.pone.0193721.g005

generate a filtered map where the different classes are homogenized enough without over-smoothing the classification result (Fig 5D and 5G).

In this study, the probability maps are obtained from the confidence scores of the SVM classification result, while the guidance image is obtained by calculating the one band representation of the HS cube by performing a dimensionality reduction using the FR-t-SNE algorithm [28].

**Unsupervised clustering segmentation.** Hierarchical clustering algorithms are able to explore the different subspaces presented in a HS cube. Each cluster centroid represents a spectra corresponding to a material in the scene, while the membership functions provide the weights for these spectra. Some works based on hyperspectral analysis for medical applications use unsupervised clustering as part of the classification algorithm, such as for colon tissue cell classification [48] or laryngeal cancer detection [49]. Unsupervised clustering provides a hierarchy of segmentations/clusters and its correspondent cluster centroid. Although it does not provide any discriminant feature by itself, it could be used delineate the boundaries of the different spectral regions presented in the HS image.

The unsupervised stage of the algorithm is based on a clustering method [50]. This method provides a segmentation map where all the different tissues, materials or substances found in the HS image are grouped forming clusters that have similar spectral characteristics. Three different clustering algorithms have been applied to the available HS images differentiating between 24 clusters: Hierarchical rank-2 non-Negative Matrix Factorization (H2NMF) [50], Hierarchical K-Means (HKM) and Hierarchical Spherical K-Means (HSKM) [51]. After a visual evaluation of the resulting maps by the specialists, it was found that all clustering methods provided useful information about the different tissues, materials and substances that were presented in the scene. Due to the fact that all three clustering methods provided similar information, HKM was selected in this study since it had the lower computational cost providing similar results. In the context of this work, the clustering process provides a good delimitation of the different areas presented in the image that should be identified by a specialist or by an automatic process, i.e., supervised classification. For this reason, a method to merge the results from the supervised and unsupervised stages of the brain cancer algorithm is required to obtain the final classification map.

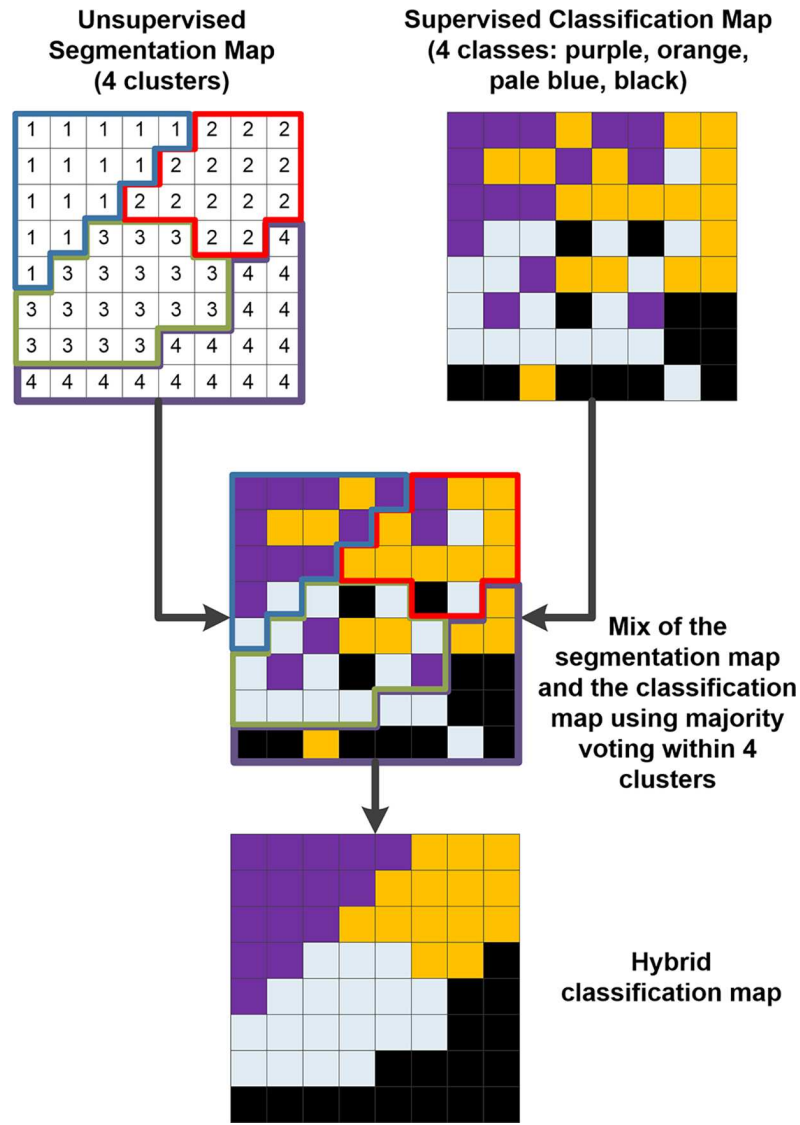

**Fig 6. Hybrid classification example based on a majority voting technique.** The unsupervised segmentation map and the supervised classification maps are merged using the majority voting method.

https://doi.org/10.1371/journal.pone.0193721.g006

**Hybrid classification.** In the previous sections, the advantages of supervised and unsupervised learning methods have been introduced. On the one hand, supervised learning can infer the knowledge previously provided by neurosurgeons and pathologists, but it can poorly provide a good delimitation of the tumor area. On the other hand, the unsupervised clustering results provide a good association of similar pixels, but each cluster is semantically meaningless. In order to solve this problem, an algorithm for merging these two sources of information has been employed. This hybrid algorithm has been previously used in hyperspectral imaging [29], and consists of a technique that merges the information from a supervised classification map and an unsupervised segmentation map (Fig 6). In the first step of this algorithm, the segmentation map and the supervised classification map are calculated independently from the same pre-processed HS cube. Once both maps have been obtained, the information is merged using the majority voting algorithm. For each cluster found by the clustering algorithm, all pixels are assigned to the most frequent class in each region in the supervised classification map. The combination of the supervised classification with the segmentation map provides some advantages. On the one hand, the unsupervised segmentation maps obtained with the clustering process have shown good capability in finding homogeneous spatial data structures from the HS cube. However, it does not provide any identification of the tissue, material or substance that the cluster belongs to. On the other hand, the supervised classification approach employs the diagnosis information provided by medical doctors (neurosurgeons and pathologists) to generate a classification map where each pixel of the image has been assigned to a certain class. However, the amount of labeled information is limited. Using the previously described MV algorithm, the strengths of each method are exploited. As stated in [29], over-segmentation (different clusters correspond to the same class) is not a crucial problem, but undersegmentation is not desired. Fig 6 graphically represents the method of the hybrid algorithm where an unsupervised map, composed by four different clusters that have no semantic meaning, is merged with a supervised classification map, composed by four different classes that have histological meaning. The final hybrid classification map represents each pixel within a certain class (identified by the supervised classification algorithm) grouped taking into account the clusters obtained by the unsupervised segmentation map (that delimitates the borders of each cluster region).

### Brain cancer detection algorithm acceleration

As far as the actual system implementation concerns, a preliminary demonstrator has been built using a modified version of the brain cancer detection algorithm [52]. To implement this application, both a computer and a hardware accelerator–MPPA-256-N, an architecture that gathers 256 processing units [53]–have been employed. On the one hand, the common stages of the application–data pre-processing and hybrid classification–and the unsupervised classification are executed on the computer that manages the hyperspectral cameras. On the other hand, the spatial-spectral supervised classification is mapped to a hardware accelerator. The rationale behind is the high computational load of the stage. For this preliminary demonstrator [52], the spatial-spectral stage has been modified. Additionally, as the system aims at building a generic classification model to assist neurosurgeons, without adding new samples, the classification model generation stage could be removed from the processing chain and consider it as a configuration step. Therefore, the processing chain would be composed of four stages: 1) a pre-processing of the HS cube; 2) a spatial-spectral supervised classification; 3) an unsupervised classification; 4) a hybrid classification.

### Evaluation metrics

The methodology for evaluating the supervised classifiers in a quantitative way is as follows: firstly, the labeled information corresponding to a simulation was used, and then, a 10-fold cross validation was applied in order to measure the performance of the model. The quantitative evaluation metrics used for this purpose are sensitivity, specificity and overall accuracy metrics. These are calculated from the following conditions:

- True Positive (TP): Correctly detected conditions. The result of the test is positive and the actual value of the classification is positive.

- False Positive (FP): Incorrectly detected conditions. The result of the test is negative and the actual value of the classification is positive.

- True Negative (TN): Correctly rejected conditions. The result of the test is negative and the actual value of the classification is negative.

- False Negative (FN): Incorrectly rejected conditions. The result of the test is positive and the actual value of the classification is negative.

*Sensitivity* is the proportion of the actual positives that are correctly identified as positives by the classifier (see Eq 3). *Specificity* is the proportion of the actual negatives that the classifier successfully valuates as negative (see Eq 4). *Overall Accuracy* refers to the ability of the model to correctly predict the class label of new or previously unseen data (see Eq 5).

$$Sensitivity = \frac{TP}{TP + FN} \qquad (3)$$

$$Specificity = \frac{TN}{TN + FP} \qquad (4)$$

$$Accuracy = \frac{TP + TN}{TP + FP + TN + FN} \qquad (5)$$

Once the quantitative metrics have been obtained, the previously trained SVM classifier is used to classify a whole HS cube, and the result is evaluated by neurosurgeons in order to analyze the quality of the algorithm in distinguishing different types of tissues, materials or substances.

## Experimental results

### Hyperspectral imaging can distinguish between tumor and normal tissue pixels by their spectra

Fig 7A and 7B show the mean and variances of the pre-processed spectral signatures of the tumor tissue, normal tissue and blood vessel labeled pixels obtained from the golden standard database of patient 1 and 2, respectively. As it can be seen in this figure, the shape of the signature depends on the tissue heterogeneity, especially in the tumor class. There are some similarities between the spectral signature of the blood vessel class and the tumor class that could

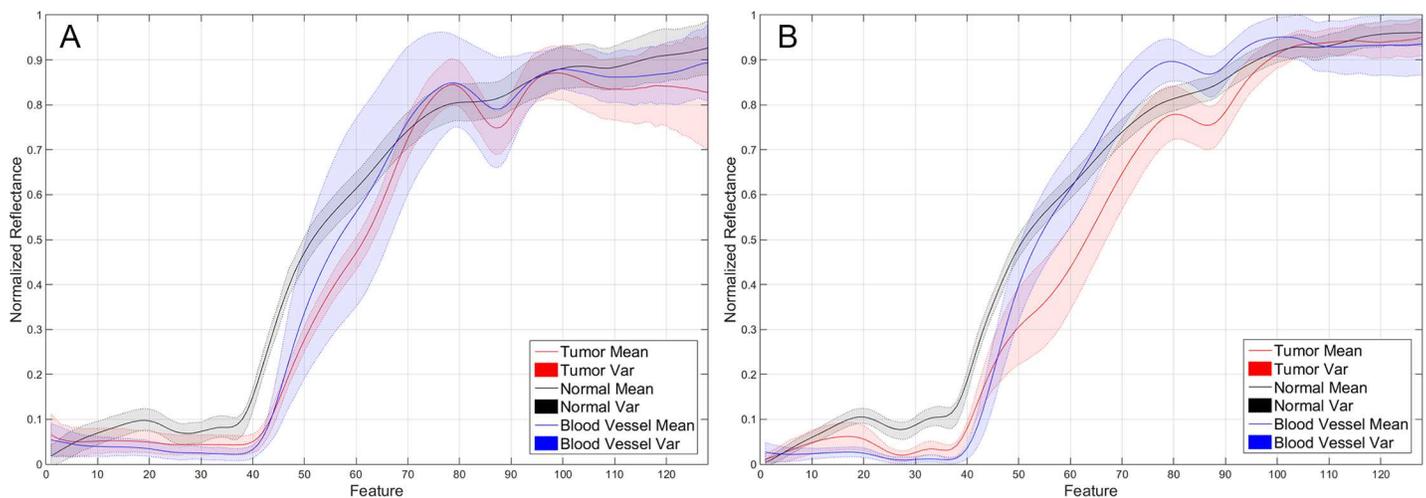

**Fig 7.** Mean and variances of the pre-processed spectral signatures of the tumor, normal and blood vessel classes of the labeled pixels from patient 1 (A) and patient 2 (B), represented in red, black and blue color respectively.

https://doi.org/10.1371/journal.pone.0193721.g007

produce some misclassifications, as it will be explained later. However, it is possible to see that the differences between the normal class and the tumor class are remarkable. These differences will ensure a successful classification of the normal and tumor pixels by the supervised classifier. In order to demonstrate that the use of a supervised classifier will achieve a reliable differentiation between the labeled pixels that conforms the golden standard database, these pixels have been spectrally analyzed employing an SVM classifier. Afterwards, the SVM model generated using the golden standard database for each patient was employed to classify the entire HS cube of this patient. As it was previously mentioned, the golden standard information was extracted from the HS data using a specific tool developed to this end.

In order to measure the supervised classifier performance and to select the optimal configuration of the SVM model, a three-way cross validation has been employed. Linear, Radial Basis Function (RBF), polynomial and Sigmoid kernels have been tested and compared. Fig 8A shows the overall accuracy classification results obtained in the experiments comparing the four SVM kernels with the default parameters, using the labeled dataset for each patient individually and performing the three-way cross validation. S1–S4 Tables present the confusion matrix results of the different classifications for each patient and type of kernel. Linear kernel provides the best accuracy results for this type of sample having a lower computational cost than the other kernels exceeding 99% of overall accuracy. This indicates that there is a strong reliability on classifying the spectral samples of the brain surface using a supervised classifier. Fig 8B and 8C illustrate the results of specificity and sensitivity metrics respectively with the linear kernel for each patient and class using the One-vs-All method. As it can be seen in these figures, the SVM classifier offers specificity and sensitivity results higher than 96%, reaching in most cases 100% specificity and sensitivity.

Fig 9A, 9B, 9C, 9D and 9E show the synthetic RGB images generated from each HS cube where the tumor area has been surrounded with a yellow line in each RGB image. Fig 9F, 9G, 9H, 9I and 9J show the golden standard maps generated using the labeling tool, where red, green, blue and black colors represent the *tumor tissue*, *normal tissue*, *blood vessels* and *background*, respectively. The qualitative results generated by the supervised classifier are shown in Fig 9K, 9L, 9M, 9N and 9O. These supervised classification maps have been obtained using the SVM model generated from the golden standard. The color representation is the same as the golden standard representation previously introduced, except for the blue color representing the hypervascularized tissue presenting on the brain surface apart from the blood vessels. In each supervised map, it is possible to identify the tumor area. Some false positives can be found in the images. This result is produced due to the spectral similarities between the tumor tissue and the main blood vessels or areas with extravasated blood in the surgical field as a result of the resection. In Fig 9K, the supervised classification map of patient 1 is shown. In this result, it can be seen that there are some false positives (delineated by an orange line) where a main blood vessel is presented (red area in the center of the image) and near other blood vessels far from the tumor area. Furthermore, there is another false positive in a small region in the right bottom of the image where the bone of the skull is visible (outside of the region where the parenchyma is exposed) due to the extravasated blood from the craniotomy. The same effect is observed with patient 3 (Fig 9M) where there are some false positives outside of the parenchymal area. Despite these false positives, the tumor area is clearly identifiable in each image, and in any case blood vessels and extra-parenchymal tissue are very evident to the surgeon during resection, so that no diagnostic confusion is likely to happen. This first step of the cancer detection algorithm results in the approximate identification of the tumor and normal tissue areas using the SVM supervised classifier. The next step is to improve the classification maps employing spatial information provided by the HS image.

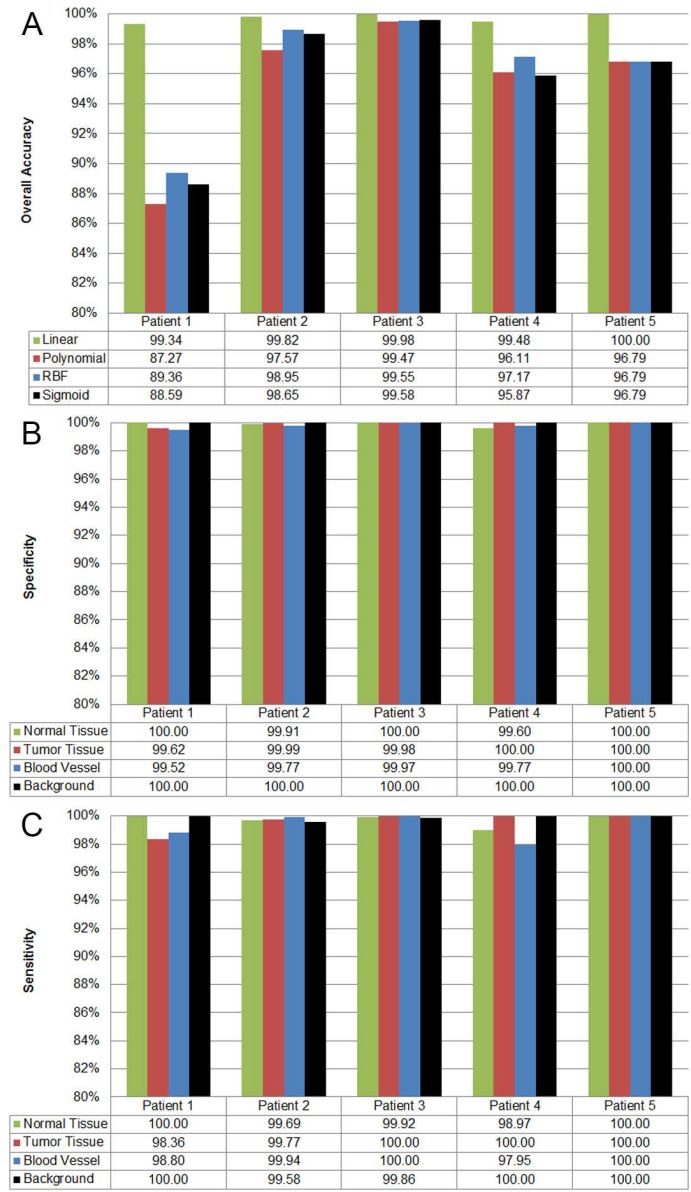

**Fig 8. Quantitative results of the supervised classification performed with the SVM classifier applied to the labeled data of each patient.** (A) Overall accuracy results of supervised classification per SVM kernel type and patient. (B) and (C) Specificity and sensitivity results obtained using the SVM classifier with linear kernel for each patient and class employing the One-vs-All evaluation method.

https://doi.org/10.1371/journal.pone.0193721.g008

### Improving the spatial coherence of the supervised classification maps

The supervised classification maps generated in the first step of the cancer detection algorithm have been improved by combining these results with a one-band representation of the HS cube using a KNN filtering method. The one-band representation of the HS image, where the most significant information of the image is revealed, has been generated using the FR-t-SNE, which offers a high contrast value compared to alternative dimensional reduction algorithms. Fig 9P, 9Q, 9R, 9S and 9T present the FR-t-SNE one-band representation of each HS cube. In these images, it is possible to identify the different areas presenting on the brain surface as

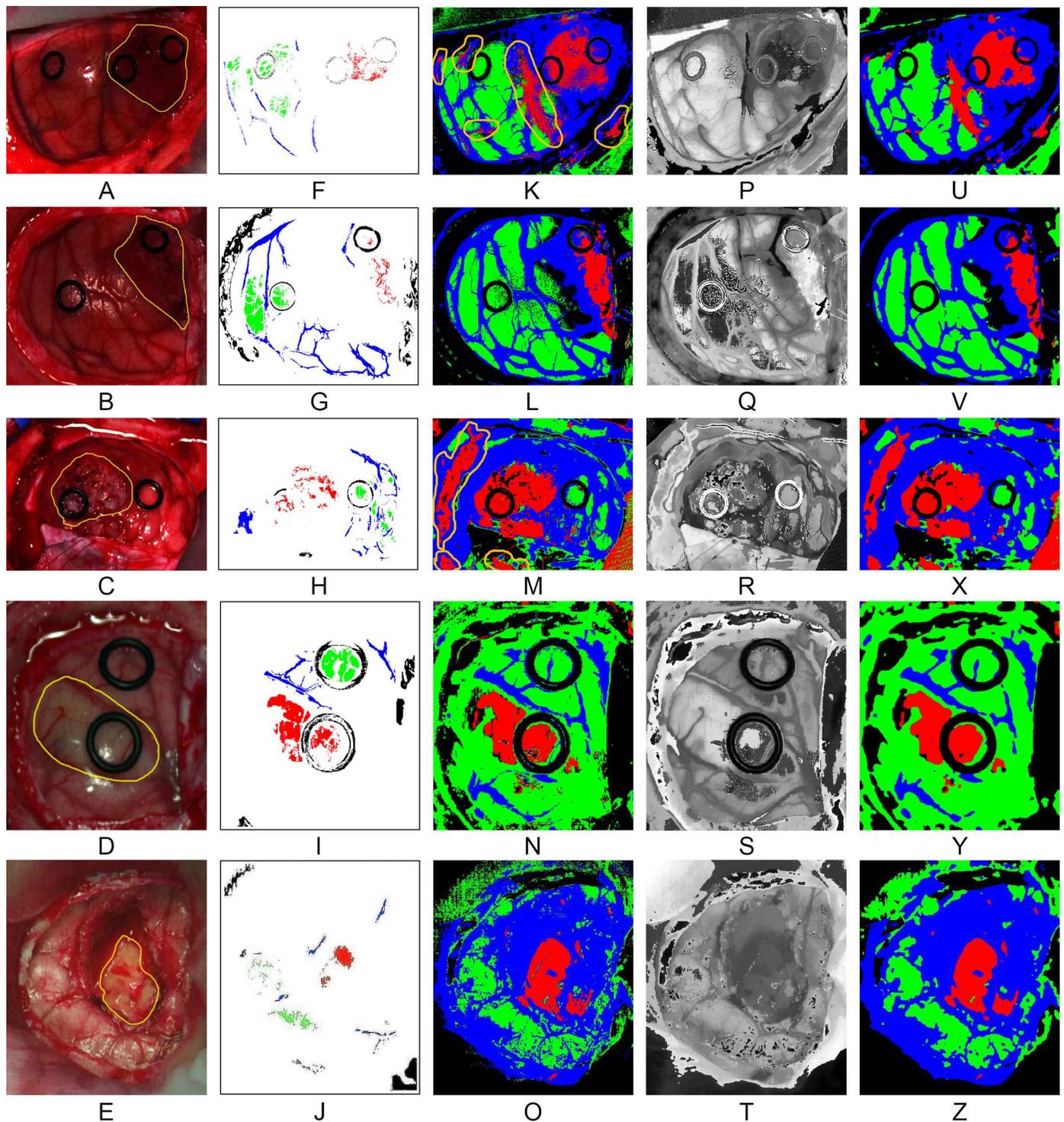

**Fig 9. Results of each step of the optimized spatial-spectral supervised classification of the five different patients.** (A), (B), (C), (D) and (E) Synthetic RGB images generated from the HS cubes. (F), (G), (H), (I) and (J) Golden standard maps used for the supervised classification training. (K), (L), (M), (N) and (O) Supervised classification maps generated using the SVM algorithm. (P), (Q), (R), (S) and (T) FR-t-SNE one band representation of the HS cubes. (U), (V), (X), (Y) and (Z) Spatially optimized classification maps obtained after the KNN filtering.

https://doi.org/10.1371/journal.pone.0193721.g009

their borders are highlighted. In these one-band representations, it is possible to identify the tumor area in each image. FR-t-SNE results together with the probability scores obtained from the supervised classification maps are the inputs for the KNN filtering. This filtering process is used to increase the spatial coherence of the supervised classification maps, providing the contextual information of each pixel in the classification scheme. Fig 9U, 9V, 9X, 9Y and 9Z illustrate the spatially optimized classification maps obtained after the KNN filtering process. It is apparent that the region of each class in the images has been homogenized giving coherence to the classification maps. Although the differences between the supervised classification maps and the spatially optimized classification maps are not very noticeable when looking at the resulting images by the naked eye (Fig 9K–9O and Fig 9U–9Z), this is a high important task since this homogenization will improve the final stage of the cancer detection algorithm, which will assign the classes to the otherwise meaningless clusters provided by the unsupervised clustering algorithm. If the number of pixels that belongs to a certain class (tumor, normal, hypervascularized or background) increases or decreases in the spatially optimized classification map, the final brain cancer classification map could be affected, showing different densities of a certain class in a certain region delimited by the unsupervised clustering algorithm.

### Unsupervised clustering for accurate boundaries delineation of the brain surface

Fig 10A, 10B, 10C, 10D and 10E show the segmentation maps generated for each patient employing the HKM clustering algorithm. As it can be seen, structures such as blood vessels, materials like the ring markers and different tissue regions are delineated by the clustering algorithm. Furthermore, the region of interest that is formed by the parenchymal area of the brain can be clearly differentiated. Inside this area, some different structures of tissue are highlighted, delimiting with high accuracy the boundaries of each region. However, the information provided by the segmentation maps is meaningless: the colors that represent each cluster are randomly selected and there is no class associated for each cluster. For this reason, it is necessary to combine the supervised identified classes with the unsupervised accurate clusters.

### Delimiting and identifying the human brain area affected by cancer

The final stage of the cancer detection algorithm has the goal of combining the segmentation map, obtained by the clustering algorithm, and the spatially homogenized classification maps, generated after the KNN filtering process, to build the final classification map employing the MV algorithm. Fig 10F, 10G, 10H, 10I and 10J show the MV classification map results. These results have been generated applying the maximum majority class of the supervised classification map to each cluster of the segmentation map. These MV maps provide more accurate results than the spatially optimized supervised classification maps. The boundaries of each class region are better delineated. In some cases, the tumor area is reduced, having mixed tissue (normal and tumor) in the area where only tumor class was presented in the supervised classification map (see patient 2, Fig 10G). The same effect is observed in patient 3, where small islands of normal tissue are found to be mixed in the tumor region (see patient 3, Fig 10H). Although this MV classification map provides better delineation of the areas affected by cancer on the brain surface, it is possible to have additional hidden information in these maps. For example, if a cluster that represents a certain class includes a zone with a high percentage (but not the maximum) of another class, this information is not revealed in the resulting image. For this reason, another visualization of the MV classification map was developed, the One Maximum Density (OMD) map. In this case, only the maximum probability results

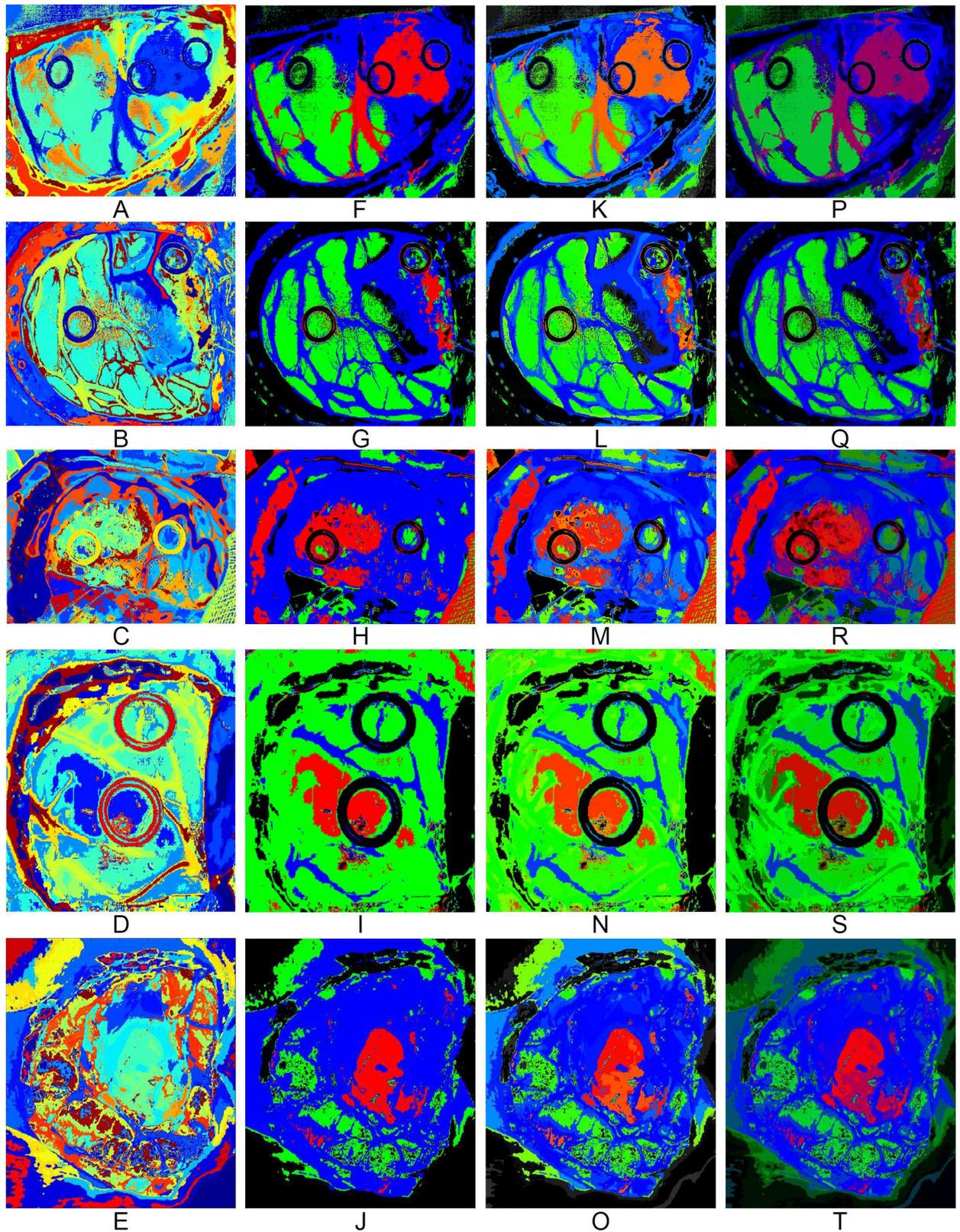

**Fig 10. Results of each step of the proposed cancer detection algorithm applied to the five different patients.** (A), (B), (C), (D) and (E) Segmentation maps generated using the HKM algorithm. (F), (G), (H), (I) and (J) MV classification maps. (K), (L), (M), (N) and (O) OMD maps

that take into account only the major probability per class obtained from the MV algorithm. (P), (Q), (R), (S) and (T) TMD maps that take into account the first three major probabilities per class obtained from the MV algorithm.

https://doi.org/10.1371/journal.pone.0193721.g010

obtained by the MV algorithm for each cluster are used to determine the color map, and the color of each class is then degraded using the percentage of the probability. For example, if the probability of the tumor class for a certain cluster is 80%, the cluster color is degraded 20% (the cluster RGB color will be R = 0.8, G = 0, B = 0). The color gradient is performed only for the tumor tissue, normal tissue and blood vessel/hypervascularized tissue classes. The background class is not degraded. Fig 10K, 10L, 10M, 10N and 10O show the OMD maps for each capture, with areas of degraded color. This observation indicates that the MV result probability was somewhat lower than for undegraded areas, and may point to the presence of different tissue classes merged in this cluster. In order to represent the classes that are mixed in a certain cluster, a third map is based on the three maximum probability values of the MV results in each cluster. This representation, the Three Maximum Density (TMD) map, offers more information from the MV results, mixing the color of each class using the percentage of the three maximum MV probability values. For instance, if the probability of tumor class for a certain cluster is 60%, the probability of normal tissue is 10% and the probability of blood vessel/hypervascularized tissue is 30%, the RGB color of the cluster will be R = 0.6, G = 0.1 and B = 0.3. By employing this technique, it is possible to visualize the clusters where their respective mixed classes are hidden. Fig 10P, 10Q, 10R, 10S and 10T show the TMD maps of each capture, where clusters that are partially mixed between the classes present darker colors. Patient 3 is a good example that contains hidden information in the MV map (Fig 10H). After the generation of the TMD map (Fig 10R), it is possible to visualize a new area surrounding the main tumor region represented in purple color, which corresponds with hypervascularized tissue with tumor infiltration. In this case, the system can estimate the proportion of malignant tissue that is mixed with the normal hypervascularized tissue. When the tissue is classified as normal (green color), there is no mixture between malignant and normal tissue. When there is some minimum amount of malignant tissue, the proportion of malignant tissue is showed in the TMD map with a gradient of red color and thus is marked for being resected in order to avoid tumor recurrence.

### Accelerating the brain cancer detection maps generation

In order to assess the application in terms of the processing time required to analyze the HS images during surgical procedures, Table 2 shows the results obtained from the five patient images employed in this research. This table presents the sequential time results obtained in a CPU implementation (using a computer with an Intel® Core™ i7-4770k 3.5GHz) and the time results obtained using the hardware acceleration in the spatial-spectral supervised classification stage. Due to the connection between the computer and the hardware accelerator, a time for the transmission is required in the accelerated version of the algorithm. However, the partition of the algorithm in both platforms allows executing the unsupervised clustering in the CPU and the spatial-spectral supervised classification in the hardware. The total time required for the processing in the accelerated version is computed taking into account the maximum time obtained between the spatial-spectral supervised classification and the unsupervised clustering. Specifically, when the hardware accelerator is not employed, the spatial-spectral supervised classification is the most time consuming stage. In contrast, an average speedup factor of 26.83x is achieved on the spatial-spectral supervised classification stage when the hardware accelerator is used. These results show that the proposed system provides a classification map of the captured scene during the surgery to neurosurgeons in approximately 1 minute, depending on the size of the captured image.

Table 2. Processing time results comparison for each patient.

| Patient ID | # Pixels | Processing Type | Pre-processing | Transmission | Spatial-Spectral Supervised Classification | Unsupervised Clustering | Hybrid Classification | Total |
|---|---|---|---|---|---|---|---|---|
| 1 | 251,532 | Sequential (s) | 14.53 | 0.00 | 482.64 | 45.44 | 0.010 | 542.62 |
| | | Accelerated (s) | | 15.10 | 16.91 | | | 75.08* |
| | | Speedup factor | 1.00 | 0.00 | 28.54 | 1.00 | 1.00 | 7.23 |
| 2 | 219,232 | Sequential (s) | 11.34 | 0.00 | 467.47 | 38.97 | 0.008 | 517.79 |
| | | Accelerated (s) | | 12.02 | 13.92 | | | 62.33* |
| | | Speedup factor | 1.00 | 0.00 | 33.57 | 1.00 | 1.00 | 8.31 |
| 3 | 185,368 | Sequential (s) | 10.28 | 0.00 | 321.26 | 33.52 | 0.008 | 365.01 |
| | | Accelerated (s) | | 11.60 | 11.98 | | | 55.35* |
| | | Speedup factor | 1.00 | 0.00 | 26.82 | 1.00 | 1.00 | 6.59 |
| 4 | 124,691 | Sequential (s) | 7.22 | 0.00 | 146.63 | 22.26 | 0.005 | 176.11 |
| | | Accelerated (s) | | 8.53 | 7.12 | | | 38.01* |
| | | Speedup factor | 1.00 | 0.00 | 20.58 | 1.00 | 1.00 | 4.63 |
| 5 | 189,744 | Sequential (s) | 14.27 | 0.00 | 268.98 | 33.93 | 0.006 | 317.18 |
| | | Accelerated (s) | | 10.53 | 10.92 | | | 58.73* |
| | | Speedup factor | 1.00 | 0.00 | 24.62 | 1.00 | 1.00 | 5.40 |

*The total time produced in the accelerated version is computed taking into account the maximum time obtained between the spatial-spectral supervised classification and the unsupervised clustering.

https://doi.org/10.1371/journal.pone.0193721.t002

## Discussion

The task to identify the boundary between the tumor tissue and the normal tissue that surrounds is difficult for neurosurgeons by only using the naked eye, as brain tumors are extremely infiltrative. The current tools employed to this end have many limitations to assist in the delineation of the tumor boundaries. MRI-based neuronavigation accuracy is affected by the brain shift during resection and depends on a questionable correlation between the extent of enhancement on MRI and cellular infiltration. Other techniques, like 5-ALA fluorescence, do not work in low-grade lesions despite being highly invasive and is not recommended for use in children. For these reasons, there is a need to develop new techniques for tumor margin delineation in real-time, maximizing the resection of the tumor and minimizing the resection of the adjacent normal brain. HSI offers a new possibility to address these issues, being a non-contact, non-ionizing and non-invasive technique.

In this research work, a methodology to develop a surgical tool for identifying and delineating the boundaries of the tumor tissue using HS images has been described. For processing these data, an active interaction between medical doctors and engineers was required. On the one hand, medical doctors generated the HS image database and the selection of the images where tumor tissue was present. They were also involved in the identification of certain types of tissues, materials and substances that appear in the captured HS cubes. On the other hand, engineers performed the digital processing of the images, developing a brain cancer detection algorithm that exploits the spatial and spectral features of the HS images.

The preliminary results obtained in the supervised classification of the tissues that have been previously labeled by the specialists, demonstrate that it is possible to accurately discriminate between normal tissue, tumor tissue, blood vessels and background with an overall accuracy higher than 99%. Using the supervised models generated with the labeled data, the entire HS images were classified and qualitatively evaluated. Five SVM classification maps obtained from five different patients affected by a grade IV glioblastoma tumor were generated. These maps can identify the regions where the tumor is located. Employing a spatial-spectral optimization method based on a KNN filtering and a FR-t-SNE dimensional reduction, the SVM classification maps were spatially homogenized. A clear identification of the tumor regions using this spatial-spectral supervised classification maps is provided. However, these maps do not offer accurate delineation of the boundaries. The unsupervised stage of the algorithm based on a HKM clustering method provides a segmentation map where the boundaries of 24 different regions with similar spectral characteristics are delineated. The fusion of the spatial-spectral supervised classification map and the unsupervised segmentation map through the MV algorithm generates the final classification map, where the boundaries of the different tissues materials or substances presented in the image are identified with a certain class. In summary, the spatial-spectral classification maps allow assigning each cluster in the segmentation map to an identifiable tissue class.

Employing the information provided by the MV algorithm, three different ways to represent the final results were analyzed. The MV map assigns the maximum probability of each class to each cluster and represent the cluster with the correspondent color: red for tumor tissue, green for normal tissue, blue for blood vessel/hypervascularized tissue and black for background. On the other hand, the OMD map displays the color of each class degraded according to the value of the first major probability. By using this technique is possible to identify the clusters that conform only slightly to their assigned class. Finally, the TMD map represents each color as a combination between the different classes mixed in a certain cluster. This map is of the most value to the operating neurosurgeon, since it offers the possibility to assess the degree of tumor infiltration into the surrounded normal brain. This assessment is key for judging the desired extent of resection.

Finally, this complex algorithm was accelerated in order to obtain the results of the classification in surgical-time during the neurosurgical operation. As a preliminary system, these results are highly promising since this acceleration allows obtaining the classification results in ~1 minute depending on the size of the image, which represents an average speedup factor of 6.43x, with respect to a sequential implementation in a CPU. Compared with the intraoperative pathological analysis or the intra-surgical magnetic resonance, that can take more than 30 minutes, we provide the classification result in ~1 minute, indicating the precise location of the tumor to the operating surgeon in surgical time.

## Limitations

The following relevant limitations have been found during this research: a) some false positives have been found on the results; b) there is a need of a clinical validation of the system; c) some misclassifications have been found between different tissues; and d) there is a need of an acceleration of the entire algorithm. Next, these limitations are detailed together with some possible solutions.

False positives have been encountered in the obtained results that could be solved with further investigations. For instance, there are some misclassifications between blood or blood vessels and tumor tissue due to the high intra-class variability between the vascularized tissues, although these false positives do not affect the area of identified tumor so that the margins of the tumor remain clearly evident. The use of an increased database to generate the supervised classification model, where the inter-patient variability is taken into account, is expected to produce better classification results. The inclusions of more labeled samples of normal tissue will reduce the occurrence of false positives in the results.

Furthermore, an extensive clinical validation is required to validate if the boundaries of the tumor area represented in the TMD map are accurately identified. Several biopsies of the boundaries of tumor area must be obtained and analyzed by the pathologists to certify the brain cancer algorithm results.

On the other hand, there are some misclassifications between different tissues with high vascularization. In some cases, extravasated blood and normal tissue affected by edema were classified as blood vessel/hypervascularized tissue (blue color). This misclassification is produced due to the similar spectral characteristics of the blood vessels and the tissue affected by edema. Further investigations where these spectral differences are included in the training of the brain cancer algorithm could alleviate this problem, or perhaps a new class could be created to identify the normal brain with high vascularization.

Finally, further investigations in the algorithm acceleration could provide the final TMD map in less than one second by using heterogeneous high performance computing system, thus obtaining real-time results. The future of this intraoperative HSI system is envisioned employing snapshot HS cameras, which can capture about ten images per second, allowing even tracking dynamic changes of the tissues.

## Conclusions

This study develops a brain cancer detection algorithm to classify HS images of brain tumor in surgical-time during neurosurgical operations. It has been demonstrated that the use of HSI as a new non-invasive surgical-time visualization tool can improve the outcomes of the undergoing patient, assisting neurosurgeons in the resection of the brain tumor. The identification of the tumor boundaries and the tumor infiltration into normal brain is highly relevant in order to avoid excessive resection of normal brain and to avoid unintentionally leaving residual tumor. Currently, further investigations are being carried out by the research team in order to generalize the results obtained, to optimize the algorithms and validate their findings, as well as to increase the image database and optimize the acquisition system. Furthermore, the use of other hardware acceleration platforms (such us GPUs or FPGAs) are currently under consideration to implement the full brain cancer detection algorithm. Such implementation must explore the design space to achieve the best tradeoff between real-time execution, memory usage and power dissipation using heterogeneous platforms. This next generation of medical HSI systems could offer neurosurgeons a real-time visualization tool to assist them during the entire process of the tumor resection providing several TMD maps per second.

## Supporting information

**S1 Table. Confusion matrix results of the SVM supervised classification with linear kernel applying the 10-fold cross validation method to each patient.**
(DOCX)

**S2 Table. Confusion matrix results of the SVM supervised classification with polynomial kernel applying the 10-fold cross validation method to each patient.**
(DOCX)

**S3 Table. Confusion matrix results of the SVM supervised classification with RBF kernel applying the 10-fold cross validation method to each patient.**
(DOCX)

**S4 Table. Confusion matrix results of the SVM supervised classification with sigmoid kernel applying the 10-fold cross validation method to each patient.**
(DOCX)

## Acknowledgments

We thank J. Morera, C. Espino, M. Márquez, D. Carrera, M. L. Plaza and R. Camacho, neurosurgeons and pathologists of the University Hospital Dr. Negrin. We also thank Paul Grundy and Victoria Wykes, neurosurgeons of the University Hospital of Southampton who helped in this research.

## Author Contributions

**Conceptualization:** Himar Fabelo, Samuel Ortega, Diederik Bulters, Gustavo M. Callicó, Adam Szolna, Juan F. Piñeiro, Guang-Zhong Yang, Bogdan Stanciulescu, Rubén Salvador, Eduardo Juárez, Roberto Sarmiento.

**Data curation:** Himar Fabelo, Samuel Ortega, Coralia Sosa, Diederik Bulters, Harry Bulstrode, Adam Szolna, Juan F. Piñeiro, Silvester Kabwama, Aruma J-O'Shanahan, Sara Bisshopp, María Hernández, Abelardo Báez.

**Formal analysis:** Daniele Ravi, B. Ravi Kiran, Diederik Bulters, Gustavo M. Callicó, Juan F. Piñeiro, Guang-Zhong Yang, Bogdan Stanciulescu, Rubén Salvador, Eduardo Juárez, Roberto Sarmiento.

**Funding acquisition:** Gustavo M. Callicó, Roberto Sarmiento.

**Investigation:** Himar Fabelo, Samuel Ortega, Daniele Ravi, B. Ravi Kiran, Diederik Bulters, Silvester Kabwama, Daniel Madroñal, Raquel Lazcano.

**Methodology:** Himar Fabelo, Samuel Ortega, Daniele Ravi, B. Ravi Kiran, Gustavo M. Callicó, Guang-Zhong Yang, Bogdan Stanciulescu.

**Project administration:** Diederik Bulters, Gustavo M. Callicó, Adam Szolna, Guang-Zhong Yang, Bogdan Stanciulescu, Eduardo Juárez, Roberto Sarmiento.

**Resources:** Diederik Bulters, Gustavo M. Callicó, Adam Szolna, Guang-Zhong Yang, Bogdan Stanciulescu, Rubén Salvador, Eduardo Juárez, Roberto Sarmiento.

**Software:** Himar Fabelo, Samuel Ortega, Daniele Ravi, B. Ravi Kiran, Daniel Madroñal, Raquel Lazcano, Abelardo Báez.

**Supervision:** Diederik Bulters, Gustavo M. Callicó, Harry Bulstrode, Adam Szolna, Juan F. Piñeiro, Guang-Zhong Yang, Bogdan Stanciulescu, Rubén Salvador, Eduardo Juárez, Roberto Sarmiento.

**Validation:** Himar Fabelo, Samuel Ortega, Daniele Ravi, Coralia Sosa, Diederik Bulters, Harry Bulstrode, Adam Szolna, Juan F. Piñeiro, Silvester Kabwama, Aruma J-O'Shanahan, Sara Bisshopp, María Hernández.

**Visualization:** Himar Fabelo, Samuel Ortega, Daniele Ravi, B. Ravi Kiran.

**Writing – original draft:** Himar Fabelo, Samuel Ortega, Daniele Ravi, B. Ravi Kiran, Coralia Sosa, Diederik Bulters, Harry Bulstrode.

**Writing – review & editing:** Himar Fabelo, Samuel Ortega, Diederik Bulters, Gustavo M. Callicó, Harry Bulstrode, Abelardo Báez, Rubén Salvador, Eduardo Juárez.